\documentclass[%
 reprint,
 amsmath,amssymb,
 aps,
,superscriptaddress]{revtex4-2}

\usepackage{graphicx}
\usepackage{dcolumn}
\usepackage{bm}
\usepackage{braket}
\usepackage[inkscapelatex=false]{svg}

\newcommand{\Plus}{\textsuperscript{+}}
\newcommand{\PlusS}{\textsuperscript{+} }
\begin{document}

\preprint{APS/123-QED}

\title{High-fidelity quantum state control of a polar molecular ion in a cryogenic environment}


\author{Dalton Chaffee}
\affiliation{National Institute of Standards and Technology, Boulder, Colorado 80305, USA}\affiliation{Department of Physics, University of Colorado, Boulder, Colorado 80309, USA.
}%
\author{Baruch Margulis}%
\affiliation{National Institute of Standards and Technology, Boulder, Colorado 80305, USA}  
\affiliation{JILA, National Institute of Science and Technology and University of Colorado, Boulder, CO, USA}%

\author{April Sheffield}
\affiliation{National Institute of Standards and Technology, Boulder, Colorado 80305, USA}\affiliation{Department of Physics, University of Colorado, Boulder, Colorado 80309, USA.
}%

\author{Julian Schmidt}
\affiliation{National Institute of Standards and Technology, Boulder, Colorado 80305, USA}\affiliation{Department of Physics, University of Colorado, Boulder, Colorado 80309, USA.
}%

\author{April Reisenfeld}
\affiliation{National Institute of Standards and Technology, Boulder, Colorado 80305, USA}\affiliation{Department of Physics, University of Colorado, Boulder, Colorado 80309, USA.
}%

\author{David R. Leibrandt}
\affiliation{Department of Physics and Astronomy, University of California, Los Angeles, California, 90095, USA.
}%
\author{Dietrich Leibfried}
\affiliation{National Institute of Standards and Technology, Boulder, Colorado 80305, USA}\affiliation{Department of Physics, University of Colorado, Boulder, Colorado 80309, USA.
}%

\author{Chin-Wen Chou}
\affiliation{National Institute of Standards and Technology, Boulder, Colorado 80305, USA}\affiliation{Department of Physics, University of Colorado, Boulder, Colorado 80309, USA.
}%

\date{\today}

\begin{abstract}
We use a quantum-logic spectroscopy (QLS) protocol to control the quantum state of a CaH\PlusS ion in a cryogenic environment, in which reduced thermal radiation extends rotational state lifetimes by an order of magnitude over those at room temperature. By repeatedly and adaptively probing the molecule, detecting the outcome of each probe via an atomic ion, and using a Bayesian update scheme to quantify confidence in the molecular state, we demonstrate state preparation and measurement (SPAM) in a single quantum state with infidelity less than $6\times10^{-3}$ and measure Rabi flopping between two states with greater than 99$\%$ contrast. The protocol does not require any molecule-specific lasers and the detection scheme is non-destructive. 

\end{abstract}

\maketitle


Quantum control of molecules is a burgeoning field with applications including precision measurement \cite{cornell_eEDM, Zelevinsky,hudson2011improved,acme2014order}, quantum information processing (QIP) \cite{Doyle_entanglement, holland2023demand, albert2020robust,Ni_gates}, and fundamental chemistry studies \cite{son2022control,Lewandowski_reactions}. Techniques for achieving such control include direct laser cooling of molecules~\cite{DeMille_SrF, Doyle_CaOCH,zhu2022functionalizing,truppe2017molecules} and formation of molecules in a desired quantum state through association of ultracold atoms~\cite{Ni_assembly, Cornish_assembly,Dieckmann_assembly}. The fidelity of molecular state control is approaching a regime relevant for QIP applications~\cite{Preskill_molQubit, Ni_gates, Ni_gates2}, with demonstrated single-state preparation, manipulation, and measurement fidelities exceeding 90\%~\cite{Ni_gates, Doyle_entanglement, Bakr_fid, Cornish_fid}, and those for a manifold of states above 99\%~\cite{Wilitsch_N2, Willitsch_N2_arxiv_2025}. 
Recently, over 99\% SPAM fidelity for a single molecular hyperfine state was demonstrated for directly laser cooled neutral molecules~\cite{holland2025demonstration}. Expanding high-fidelity quantum control capabilities to a wider variety of molecular species such as molecular ions will unlock molecules' full utility across quantum science applications.

Quantum-logic spectroscopy (QLS) in an ion trap is an established and general technique \cite{QLS} that has been used to gain control over ions for which standard state preparation and measurement (SPAM) techniques that rely on repeatedly scattering photons are infeasible.  Briefly, a ``logic'' ion, amenable to laser cooling and easy to control, is co-trapped with a ``spectroscopy'' ion that lacks an accessible cycling transition for cooling and fluorescence detection. Because the translational motion of the two ions within the trap is coupled, the spectroscopy ion can be sympathetically cooled by the logic ion. By mapping the internal state of the spectroscopy ion to the shared motion, information about the spectroscopy ion can be transferred to, and read out from, the logic ion. QLS, which boasts broad applicability due to its minimal requirements on the spectroscopy ion's internal structure, has recently been applied to molecular (spectroscopy) ions with promising results in precision measurement \cite{Schmidt_MgH, Chou_prep+manip, Wilitsch_N2, holzapfel2025quantum} and QIP \cite{lin2020quantum, kenBrown2024dpql}. The practicality of molecules in QIP would be further validated by achieving high-fidelity SPAM. 

Rotational transitions in polar molecules driven by ambient thermal radiation (TR) can limit state lifetimes and SPAM fidelity in molecular quantum information experiments. For a linear molecule, rotational states are characterized by the rotational quantum number $J = 0,1,2,\ldots$, with energy level spacing approximately given by  $h\nu_{J,J+1} \approx 2hB_R(J+1)$, where $B_R$ is the molecule's rotational constant. The rate at which TR-induced transitions are driven is proportional to the ambient photon energy density. For an ideal blackbody environment this density is described by Planck's law, leading to a stimulated transition rate that decreases superlinearly with decreasing temperature. Additionally, the number of thermally occupied rotational levels---calculated from the level spacing and Boltzmann statistics---scales approximately with $\sqrt{T}$. Cold environmental temperatures also reduce ambient pressure, suppressing background gas collisions. Operating in a cryogenic environment is thus highly advantageous for controlling molecular quantum states.

In this Letter, we use QLS to control a CaH\PlusS ion in a cryogenic environment and achieve high SPAM fidelity. We observe rotational state lifetimes of $18\pm2 (10\pm1)$~s for states with $J=1 (2)$, reducing the dominant error mechanism in our system by an order of magnitude over a comparable room-temperature apparatus~\cite{Chou_prep+manip,Chou_tracking}. Operating in a cryogenic environment dramatically reduces the size of the thermally populated molecular state space, with population found in $J\in\{1,2,3\}$ 97\% of the time. By adaptively probing the molecule multiple times, detecting the outcome of each probe via a QLS detection protocol, and using a Bayesian probability inference scheme to quantify our confidence in the populated molecular state~\cite{Hume_Bayes,erickson2022high}, we demonstrate SPAM in a single quantum state with $<6\times10^{-3}$ infidelity. Our protocol does not require any molecule-specific lasers and our detection scheme is non-destructive. 

\begin{figure}[]
\includegraphics[width=0.48\textwidth]{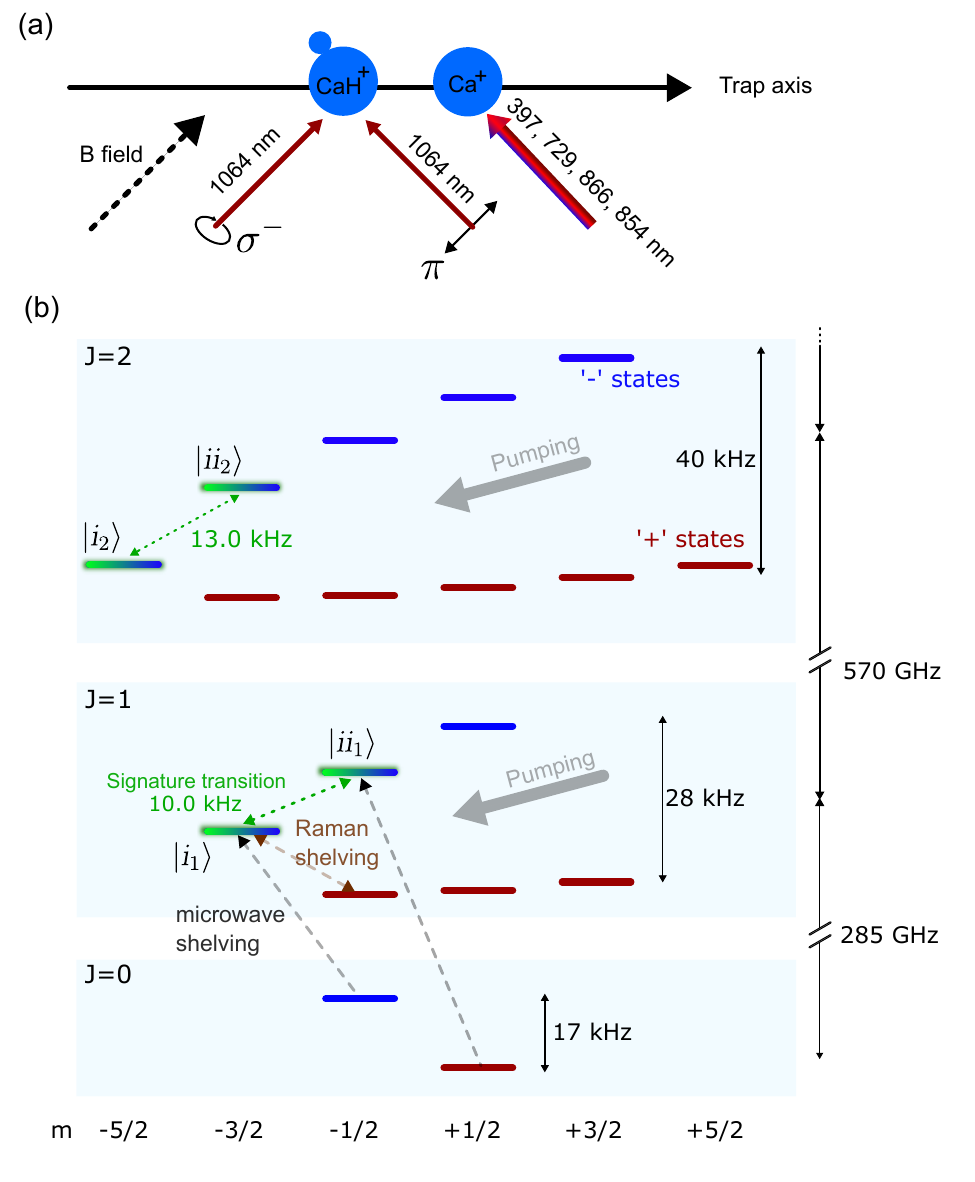}

\caption{\label{fig:apparatus} 
(a) Schematic of the two-ion crystal, static magnetic field (quantization axis, 0.4 mT), and laser beam geometry. Molecular Raman transitions are driven by two 1064~nm beams. Doppler and sideband laser cooling of the two-ion crystal is performed on the atomic ion. (b)~CaH\PlusS level structure. The molecule primarily occupies the $J=0-2$ rotational manifolds and, in equilibrium, is spread over all associated spin-rotational Zeeman sublevels. While the molecular state is unknown, Raman and microwave transitions are used to pump the molecule to $|\textsl{i}_J\rangle$, $J\in\{1,2,3\}$. High-fidelity preparation (in $|\textsl{i}_J\rangle$) and measurement is accomplished by probing back and forth with motion-adding sidebands on the signature transition, $|\textsl{i}_J\rangle\leftrightarrow|\textsl{ii}_J\rangle$, and evaluating the result of each probe via a QLS protocol. A Raman shelving pulse that drives transitions between the prepared state ($|\textsl{i}_1\rangle$) and an auxiliary level ($\ket{1, -1/2, +}$) may be introduced between preparation and measurement.}

\end{figure}

%
The experimental setup used in this work is detailed in the Supplementary Information section. In short, a \textsuperscript{40}Ca\Plus$-$\textsuperscript{40}CaH\PlusS ion crystal is trapped within a linear Paul trap~\cite{wineland1998bible,brewer2019al+} in a cryogenic environment. The trap sits inside two nested radiation shields. The inner shield reaches 15.8~K under normal operating conditions. A schematic of the ion crystal, the static magnetic field, and the relevant laser beam directions is shown in Fig.~\ref{fig:apparatus}a. A 0.40~mT magnetic field provides a quantization axis. The Ca\Plus~$S_{1/2}\leftrightarrow P_{1/2}$ transition at 397~nm is used for both Doppler cooling and fluorescence detection, with repumping via the $D_{3/2}\leftrightarrow P_{1/2}$ transition at 866~nm. Resolved sideband cooling of several motional modes is performed on the Ca\Plus~$S_{1/2}\leftrightarrow D_{5/2}$ transition  at 729~nm, with repumping via the $D_{5/2}\leftrightarrow P_{3/2}$ transition at 854~nm \cite{Leibfried_CSBC}. The most critical motional mode to cool is the axial, out-of-phase (OOP) mode at $\nu_m=4.4$ MHz, which is used to map information from the molecule to the atom during the QLS sequence. Using a projective purification step as described in Ref.~\cite{Chou_prep+manip}, the crystal can be prepared in the motional ground state $\vert{n=0}\rangle$ of the axial OOP mode with $>$99\% probability.


In equilibrium, CaH\PlusS can be described by a distribution of rotational state probabilities within the $1^1\Sigma$ vibronic ground state. At cryrogenic temperatures, rotational transitions with $\Delta J = \pm1$ occur on an $\mathcal{O}$(10~s) timescale; these are primarily due to TR, with some contribution from spontaneous decay. Each CaH\PlusS rotational manifold consists of $4J+2$ sublevels with the same $J$ but different orientations of the proton and rotational magnetic moments relative to each other and the quantization axis. The sublevels are spaced by $\mathcal{O}$(10~kHz) due to Zeeman splittings and spin-rotation interactions; details of this structure and the associated pumping strategy used in this work have been described previously~\cite{Chou_prep+manip, Chou_tracking}. The molecular sublevels $\ket{J,m,\xi}$ are characterized by the following quantum numbers: $J$, the rotational quantum number; $I=1/2$, the spin of the hydrogen nucleus; $m=m_I+m_J$, the sum of the projection quantum numbers $m_I$ and $m_J$ of nuclear spin and rotational angular momentum, respectively; and $\xi$, which distinguishes two possible superpositions of product states $\ket{J,m_J}\ket{I,m_I}$ (with the same total $m$ value but opposite $m_I$) by denoting the relative sign of the superposition. For states with $m=\pm (J+I)$, $\xi$ is defined to be the sign of $m$.


We identify a “signature manifold" consisting of two sublevels within each rotational manifold $J\in\{1,2,3\}$. We label the manifold sublevels as $\ket{\textsl{i}_J}\equiv\ket{J,-J-1/2,-}$ and $\ket{\textsl{ii}_J}\equiv\ket{J,-J+1/2,-}$. The “signature transition" $\ket{\textsl{i}_J} \leftrightarrow \ket{\textsl{ii}_J}$ has a unique frequency $\nu_J$ for each rotational manifold within the vibronic ground state. We drive transitions between sublevels within a rotational manifold using two Raman beams at 1064~nm with $\pi$ and $\sigma^-$ polarization, far-detuned from any intermediate state. By setting the frequency difference to either $\nu_J$ or $\nu_J\pm\nu_m$, we address carrier or motional sideband transitions, respectively. A pumping sequence concentrates the molecular population into $\ket{\textsl{i}_J}$ for $J\in\{1,2,3\}$ when the state is unknown, which includes a 285 GHz microwave drive to deshelve out of the long-lived $J=0$ states to sublevels in the $J = 1$ manifold~\cite{Chou_tracking}. The sublevel diagram of $J\in\{0,1,2\}$ and associated experimental operations are shown in Fig.~\ref{fig:apparatus}b.

We use a QLS protocol to determine whether the molecule occupies a particular signature manifold~\cite{Chou_prep+manip}. The sequence begins by preparing Ca\PlusS and the axial OOP mode in the $\ket{D_{5/2}}\ket{n=0}$ state, where $n$ is the motional quantum number of the axial OOP mode. The Raman beams are used to probe either the $\ket{\textsl{i}_J}\ket{n=0}\rightarrow\ket{\textsl{ii}_J}\ket{n=1}$ or $\ket{\textsl{ii}_J}\ket{n=0}\rightarrow\ket{\textsl{i}_J}\ket{n=1}$ motional sideband transition. After the probe pulse, a $\ket{D_{5/2}}\ket{n} \rightarrow \ket{S_{1/2}}\ket{n-1}$ sideband pulse is applied to Ca\Plus, which will only change the atom’s electronic state if the molecular probe has added a motional quantum to the axial OOP mode. A subsequent fluorescence detection differentiates the Ca\PlusS electronic states, thus determining whether a molecular transition occurred. A fluorescing Ca\PlusS ion ideally projects the molecule into the final state of the addressed transition.

In our experiment, a single signature transition sideband probe has $\sim85\%$ population transfer efficiency. To achieve $\gtrsim85\%$ detection fidelity, we attempt multiple, non-destructive signature transition probes. Based on the binary result $R \in \{S,D\}$ (detection of Ca\PlusS in $\ket{S_{1/2}}$  or $\ket{D_{5/2}}$ after the QLS sequence), the confidence $P(\text{in})$ that the molecules lies within the probed signature manifold is updated according to Bayes' rule~\cite{Hume_Bayes}. Specifically, given an initial probability $P(\text{in})$ that the molecule occupies the signature manifold and an outcome $R$ of a QLS sequence, the confidence is updated according to:

\begin{equation}\label{eqn:bayes}
    P(\text{in}|R)=\frac{P(R|\text{in})P(\text{in})}{P(R|\text{in})P(\text{in})+P(R|\text{out})(1-P(\text{in}))}
\end{equation}
We empirically estimate $P(S|\text{in}) = 0.85 = 1 - P(D|\text{in})$, corresponding to the observed single-molecular-probe success probability after high-fidelity Bayesian state preparation, and $P(S|\text{out}) = 0.009 = 1 - P(D|\text{out})$, corresponding to the purification level of $\ket{D_{5/2}}\ket{n=0}$ prior to the molecular probe. A predetermined confidence threshold $C_T$ is selected, and $P(\text{in})$ is initialized to 0.5. For a detection sequence, probes are repeated until one of the following 3 conditions are met:

\begin{itemize}
    \item $P(\text{in}) > C_T$ AND the final detection with $R=S$ is on the $\ket{\textsl{ii}_J}\ket{n=0}\rightarrow\ket{\textsl{i}_J}\ket{n=1}$ transition (in-manifold detection) 
    \item $1-P(\text{in}) > C_T$ (out-of-manifold detection)
    \item 25 probes have been attempted
\end{itemize}

In general, the experimental sequence is composed of “preparation", “experiment", and “measurement" stages. Preparation in $\ket{\textsl{i}_J}$ consists of making an in-manifold detection (possibly including pumping steps if the molecular state is initially unknown); note the directionality requirement for the final probe of an in-manifold detection means that the single quantum state $\ket{\textsl{i}_J}$ will be prepared with high probability. Measurement consists of a detection attempt in the prepared manifold to determine whether or not the molecule still resides there. One or several experiment pulses may be applied between preparation and measurement. A detailed description of the experimental sequence, including probe logic and Bayesian probability inference, is presented in the End Matter.

\begin{figure}[]
\includegraphics[width=0.47\textwidth]{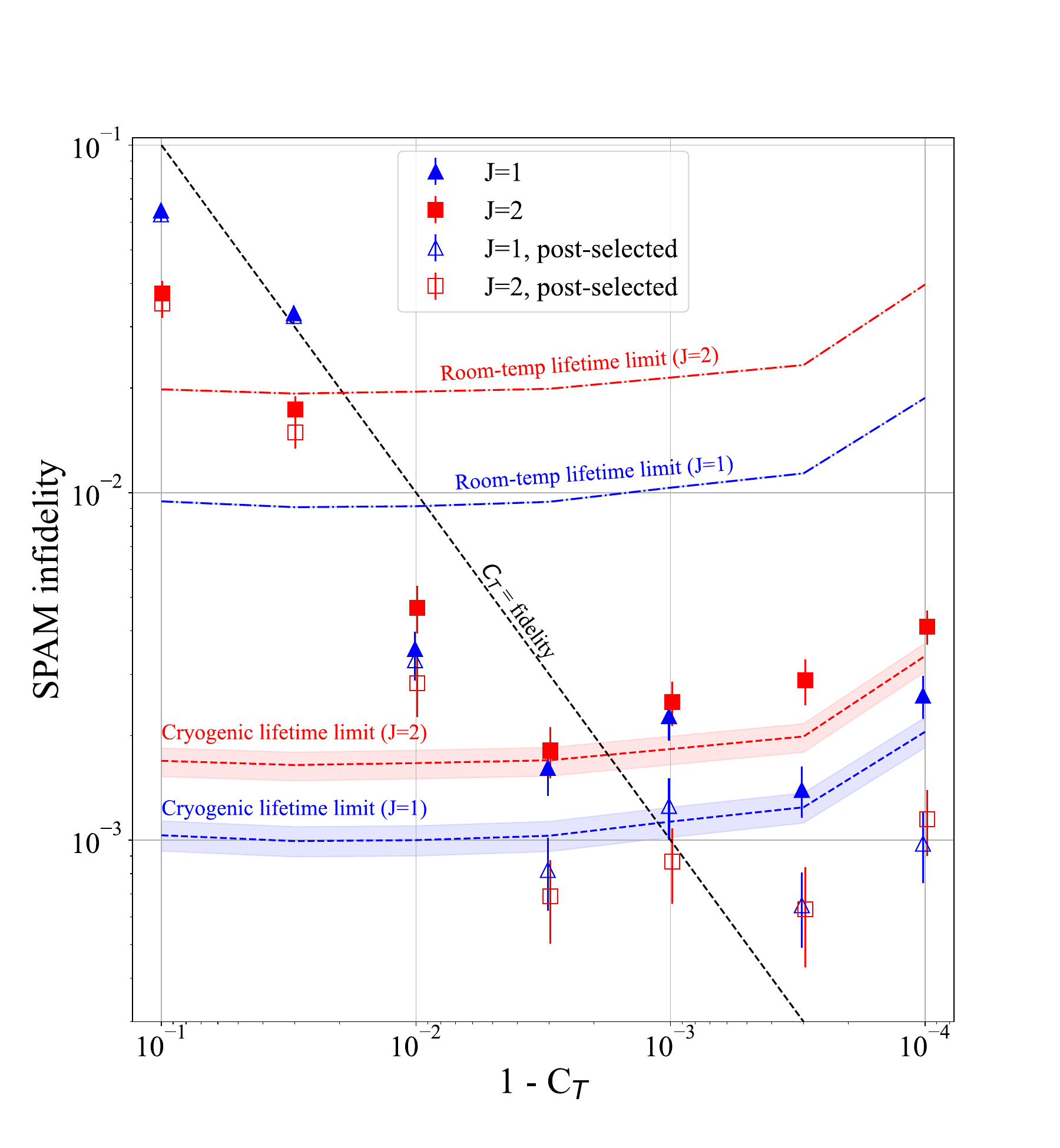}
\caption{\label{fig:hiFid} Results of SPAM infidelity characterization vs confidence threshold $C_T$ for the $J=1$ (blue triangles) and $J=2$ (red squares) signature manifolds. Infidelities are calculated by observing the fraction of measurement sequences that determine the molecule to be out of the prepared manifold immediately after preparation. Post-selected infidelity omits measurements that are followed by preparation in a different rotational level. Error bars denote one Wilson interval. For $C_T \lesssim 0.997$, infidelity tracks $1-C_T$ (black dashed line). Lifetime-limited infidelity (blue (red) dashed curves for $J=1(2)$) is calculated by dividing the observed average measurement sequence duration by the observed average lifetime for each data point; the uncertainties, indicated by the shaded backgrounds, are chiefly due to uncertainties in the experimentally-determined lifetimes. As $C_T$ approaches 1, more and more probes are needed for measurement, increasing the likelihood of population loss during measurement. For comparison, we also provide the expected limit in a room-temperature environment (dot-dashed lines) based on theoretical lifetimes.}
\end{figure}

%
%
To characterize SPAM fidelity within a given signature manifold, we omit the experiment stage. We define SPAM infidelity as the probability that the molecule is measured to be out of the prepared manifold after preparation. Fig.~\ref{fig:hiFid} presents the measured SPAM infidelities for the $J=1$ and 2 signature manifolds as a function of $C_T$.

%

Varying $C_T$ is a tradeoff between confidence achieved and number of probes required per detection event. In the absence of other error mechanisms, the measured fidelity is expected to be at or above $C_T$ (black dashed curve); this is true for our data for $C_T \leq0.997$. For $C_T >0.997$, other error mechanisms contribute significantly. For the $C_T$ with the lowest infidelity, we observe an infidelity of $(1.39\pm 0.23)\times 10^{-3}$ for $J=1$ ($C_T=0.9997$, with an average of 2.96 probes taking 22.5 ms per measurement) and $(1.81\pm 0.30)\times 10^{-3}$ for $J=2$ ($C_T=0.997$, with an average of 2.37 probes taking 17.3 ms per measurement). Reported uncertainties are statistical and represent one standard deviation.

Despite the reduction in TR due to cryogenic operation, rotational level changes during measurement are the leading error mechanism at high $C_T$. To quantify the expected error due to rotational state change for a given data point, we compare the average duration of a measurement sequence to the observed rotational manifold lifetime. The average number of probes per measurement ranges from 2.3 to 4.9 across the range of $C_T$'s considered here, the average probe duration is 7.6 (7.3) ms for $J=1(2)$, and the observed rotational state lifetimes are 18$\pm2$ (10$\pm1$) s, as detailed below. The lifetime-limited fidelities plotted in Fig.~\ref{fig:hiFid} (colored, dashed curves) are calculated by dividing the average measurement duration by the observed lifetime, representing the fraction of measurements during which a rotational state change is expected to occur. These limits provide an explanation for the increasing infidelities at high $C_T$ (since more probes, and thus more time, are required to achieve the required confidence) and the higher infidelities for the shorter-lived $J=2$ manifold, compared to $J=1$, at high $C_T$.

To quantify contribution from other error mechanisms, we can post-select away measurements that are immediately followed by preparation in a different rotational level. By doing so, we omit instances where a rotational transition has occurred during measurement. The post-selected infidelity is as low as $(6.5\pm1.6)\times 10^{-4}$ for $J=1$ and $(6.3\pm2.0)\times 10^{-4}$ for $J=2$, both for $C_T = 0.9997$. Sources for this residual infidelity may include collisional ion reordering events and fluctuations in the signature transition sideband frequency. Further discussion is provided in the End Matter.

At 295 K, the theoretical rotational level lifetimes are $2.0512\pm 0.0008$ $(0.8825 \pm 0.0015)$ s for $J=1(2)$ where the uncertainty in the theoretical lifetimes originated from the uncertainty in the permanent dipole moment of the molecule \cite{collopy2023effects}. Both theoretical and experimentally measured~\cite{Chou_tracking} lifetimes are about an order of magnitude shorter than what is observed in the cryogenic environment, corresponding to an elevated theoretical fidelity limit. These limits are shown by the dot-dashed curves in Fig.~\ref{fig:hiFid}, which are calculated according to the theoretical rotational level lifetimes and observed average measurement durations. From these limits, we infer that a room-temperature system using the same SPAM scheme would be limited to infidelities $\gtrsim0.01$. Thus, the reduced ambient TR enables improvement of SPAM infidelity by about an order of magnitude. 

\begin{figure}[htb!]
\includegraphics[width=0.49\textwidth]{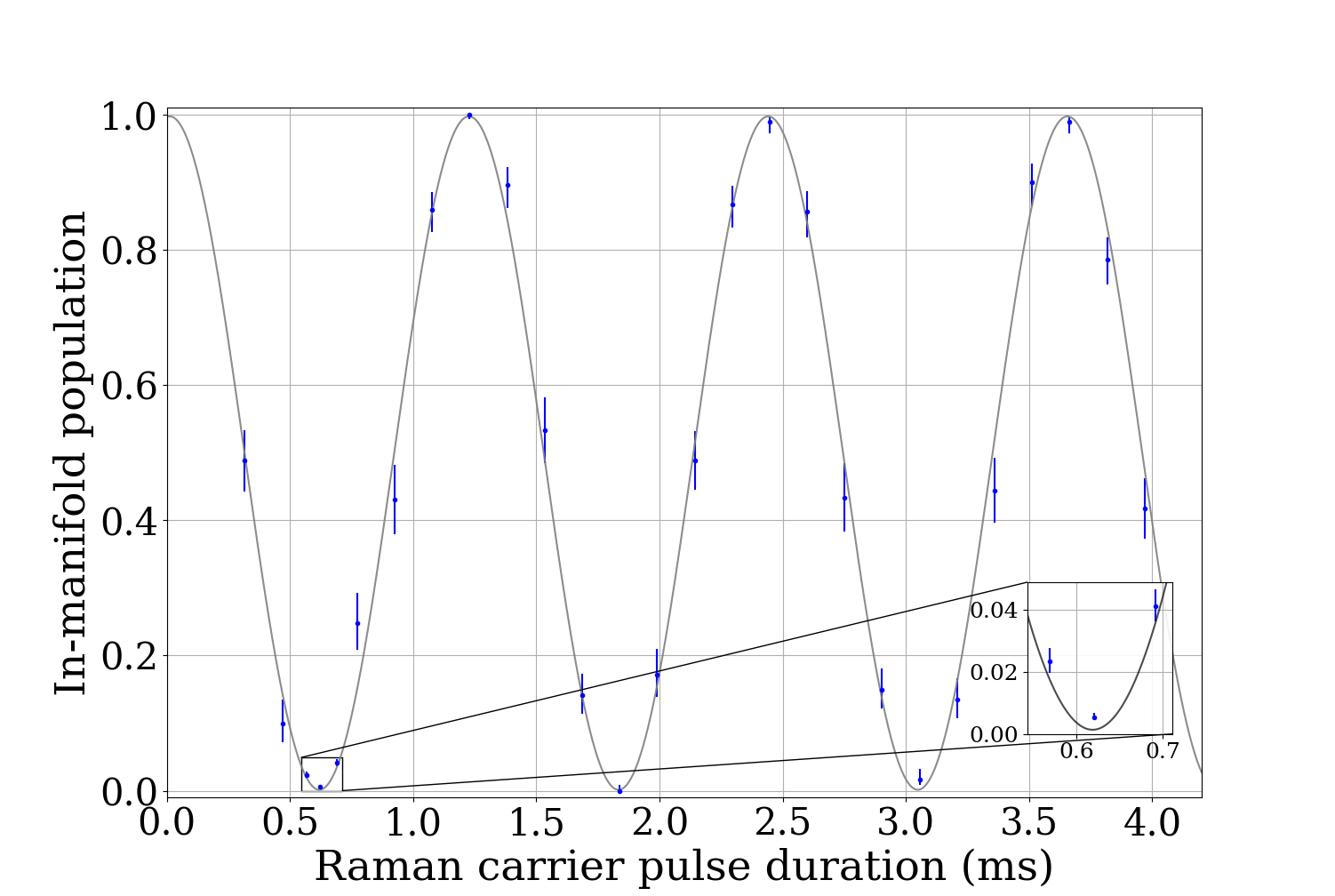}
\caption{\label{fig:carrier} Rabi flopping on $\ket{\textsl{i}_1} \leftrightarrow\ket{1,-1/2,+}$ by a pulse of variable duration that is resonant with the transition and applied between preparation and measurement. The fit is to an exponentially-decaying sinusoidal function.
After a $\pi$-pulse (620 $\mu$s duration), the molecule is measured in manifold with probability $(5.6\pm1.2)\times10^{-3}$. Error bars denote one Wilson interval.}
\end{figure}

Since our measurement does not distinguish between states within the signature manifold, the SPAM fidelity results presented thus far cannot be considered to be of a single quantum state. Nevertheless, we expect that our preparation sequence concludes with the molecule in state $\ket{\textsl{i}_J}$ with near-unity probability due to the requirement that preparation ends with a $\ket{\textsl{ii}_J}\rightarrow\ket{\textsl{i}_J}$ probe. As proof of the purity of the prepared state and further demonstration of our molecular quantum state control capabilities, we introduce an experimental pulse with variable duration coherently transferring population from the prepared sublevel $\ket{\textsl{i}_1}$ to the out-of-manifold sublevel $\ket{1, -1/2, +}$ via a Raman carrier drive. The resulting Rabi flopping, measured with $C_T$ set to 0.997, is shown in Fig.~\ref{fig:carrier}. Additional data was taken near the $\pi$-time for better statistics; after a $\pi$-pulse, population was measured to be in manifold with $(5.6\pm1.2)\times10^{-3}$ probability. We take this number as an upper bound on our single-quantum-state SPAM infidelity, though it includes some contribution from imperfect state transfer.
%
%

By continuously attempting to prepare and measure the molecular state for $J\in\{1,2,3\}$, we actively track the ion’s rotational state occupation and infer rotational level lifetimes. Doing so, we can extract information about the TR environment interacting with the ion, assuming rotational transitions are driven solely by TR~\cite{Chou_tracking}. The population fractions, ion lifetimes, and inferred blackbody temperature for each rotational state $J\in\{1,2,3\}$ are shown in Table~\ref{tab:table1}. These data were obtained by tracking the molecular rotational level for about 1 hour by running SPAM experiments with $C_T$ set to 0.997. Note that the ion was actively deshelved out of $J=0$ if suspected to reside there. The blackbody radiation (BBR) temperature is inferred from the lifetime assuming an ideal blackbody environment.



\begin{table}[htbp]
\caption{\label{tab:table1}%
Rotational level tracking over 64 minutes of repeated preparation and measurement attempts with active deshelving out of $J=0$ states. The lifetime is found by dividing the total time spent within a level by the number of times the molecule was observed to leave the state, and the effective BBR temperature is found by comparing the observed lifetime to theory~\cite{Chou_tracking,koelemeij2007blackbody}. Uncertainties in measured lifetimes are statistical, while uncertainties in inferred BBR temperature include both statistical contributions and uncertainty in the molecule's permanent dipole moment~\cite{collopy2023effects}. We cannot directly measure population in $J = 0$, but we assume the molecule has transitioned there if it is detected in $J = 1$ only after a $J = 0\rightarrow1$ deshelving pulse. The unaccounted-for 0.03 population fraction is attributed primarily to occupation of $J>3$, which we do not attempt to detect.
}
\renewcommand{\arraystretch}{1.25}
\begin{ruledtabular}
\begin{tabular}{cccc}
\textrm{$J$}&
\textrm{Occupation fraction}&
\textrm{Lifetime (s)}&
\textrm{BBR temp. (K)}\\
\colrule
1 & 0.48& $18.1\pm1.8$& $42.0^{+4.3}_{-3.7}$\\
2 & 0.36& $10.2\pm0.9$& $33.6^{+3.2}_{-2.9}$\\
3 & 0.13& $5.5\pm0.6$& $30.6^{+4.1}_{-3.6}$\\
\end{tabular}
\end{ruledtabular}
\end{table}

The inferred photon densities are significantly higher than what would be expected from thermal equilibrium with the measured inner shield temperature of 15.8 K. Moreover, considering the occupation fractions of the various $J$ levels, we find that the distribution is most consistent with a BBR temperature of 34 K~\cite{patel2025precise}. We have ruled out elevated trap RF electrode temperature, TR leakage and background gas collisions as possible explanations for this observation. Additional hypotheses include Laser beams induced effects and geometrical effects. Further discussion can be found in the Supplementary Information. Accurate identification and characterization of the leading mechanism responsible for the observed discrepancy is beyond the scope of this paper and is left for future investigation.

%
%

In summary, we have demonstrated high-fidelity molecular quantum state control, achieving better than $99.8\%$ SPAM fidelity within signature manifolds and above $99.4\%$ quantum state transfer, setting a lower bound on our single-state SPAM fidelity. We also measure enhanced rotational state lifetimes due to reduced thermal radiation. The versatile QLS protocol in a cold thermal environment constitutes a pristine setting for studying numerous molecular ion species. Species with long state lifetimes, such as non-polar ones, can be probed repeatedly and controlled with high fidelity. Moving forward, we aim to achieve fully deterministic control of the molecular quantum state and extend our experiments to other molecular species.  To these ends, we have installed a molecular beam aimed at the trapping region to be used for loading a broad variety of molecules and plan to drive rotational Raman transitions using a frequency comb~\cite{Chou_comb}. The high-fidelity quantum state control and generalizability of our apparatus promises to enable a new regime of precision measurement, QIP, and chemistry applications with molecules.

We thank D. B. Hume and A. Kwiatkowski for their careful reading of the manuscript. This work is supported by the Army Research Office, agreement number: 22X027, and the Air Force Office of Scientific Research, grant number: FA9550-23-1-0028. B.M. acknowledges the support of the Rothschild Foundation.
\newpage

\section{End Matter}

\subsection{Bayesian detection sequence}

\begin{figure}
    \centering
    \includegraphics[width=0.96\linewidth]{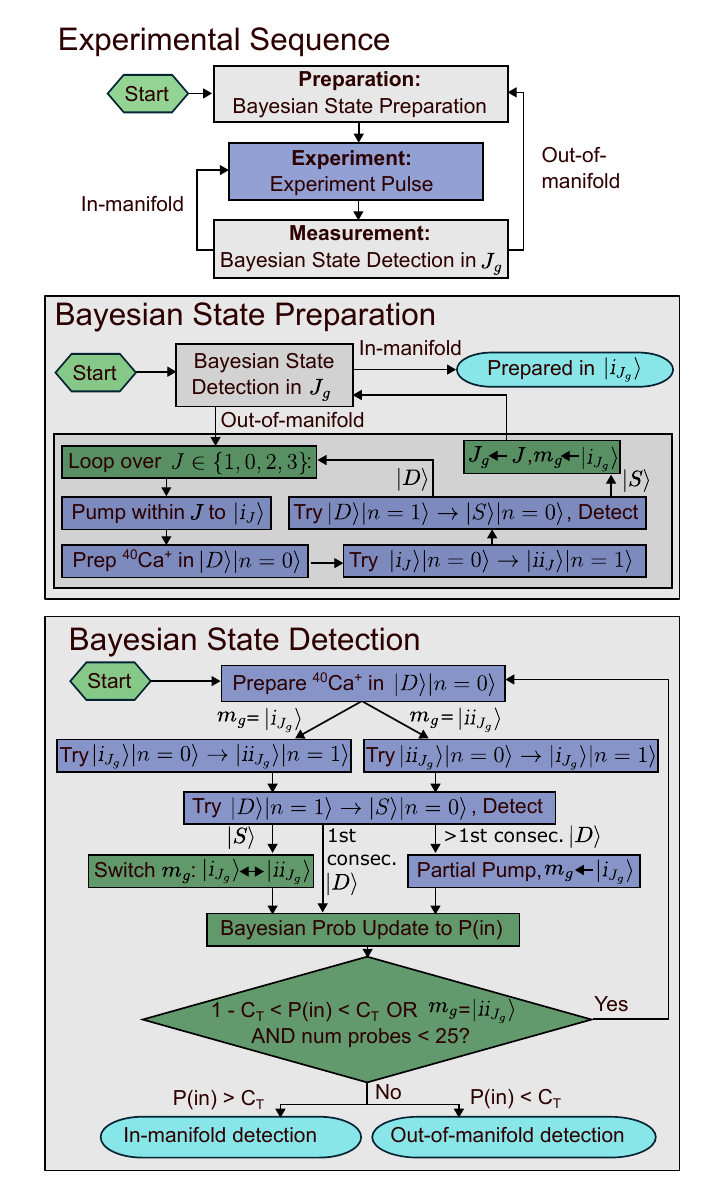}
    \caption{Full experimental sequence, including Bayesian state preparation and detection. Here, $J_g$ and $\ket{m_g}$ are the system's current determined molecular rotational and signature manifold state, respectively: they are initialized to $J_g=1$ and $\ket{m_g}=\ket{\textsl{i}_J}$. Grey boxes represent sub-protocols expanded elsewhere in the flowchart, dark blue boxes indicate physical actions performed on the ions, green boxes indicate software logic updates, and light blue ovals indicate the endpoint of a subprotocol. The experimental sequence is performed (with occasional breaks for ion order check and servo experiments) until the desired amount of data has been collected.}
    \label{fig:condscheme}
\end{figure}

A flowchart representing the experimental sequence, including the preparation, experiment, and measurement stages, is presented in Fig.~\ref{fig:condscheme}. Preparation may involve many pumping pulses if the initial state is unknown. The ``pump within $J$ to $\ket{\textsl{i}_J}$'' action consists of multiple rounds of cooling to the motional ground state and then driving directional, motion-adding sidebands on molecular transitions to move molecular population from any initial sublevel to $\ket{\textsl{i}_J}$. Due to molecular selection rules, the $J=0$ transition cannot be driven by our 1064 nm Raman beams. To detect the molecule when it is initially in the $J=0$ manifold, we deshelve population from $J=0$ into $\ket{\textsl{i}_1}$ by executing $\ket{J=0, m=-1/2,\xi=-} \leftrightarrow \ket{\textsl{i}_1}$ and $\ket{J=0, m=1/2,\xi=+} \leftrightarrow \ket{\textsl{ii}_1}$ deshelving pulses using a 285 GHz microwave source. A subsequent successful preparation in $J=1$ will indicate that the molecule was originally in $J=0$, but prepared in $\ket{\textsl{i}_1}$.

``Partial pump'' refers to a pumping pulse on the $\ket{\textsl{ii}_J} \rightarrow \ket{\textsl{i}_J}$ transition (and in $J=1$, a preceding $\ket{\textsl{iii}_1}\equiv\ket{J=1,m=+1/2,\xi=-}\rightarrow\ket{\textsl{ii}_1}$ pumping pulse). This action is necessary due to a small probability to transfer the molecular state without an $\ket{S}$ detection (as well as the possibility of off-resonant excitation from $\ket{\textsl{ii}_1}\rightarrow\ket{\textsl{iii}_1}$). Because of partial pumping, the Bayesian detection sequence does not strictly detect one particular initial sublevel, but rather either sublevel in the signature manifold (and for $J=1$, including $\ket{\textsl{iii}_1}$).

The recorded result of a measurement typically corresponds to the result (``in-manifold'' or ``out-of-manifold'') of its Bayesian state detection sequence. However, in the rare case where the detection sequence terminates due to hitting the maximum allowed number of probes (and thus with $1 - C_T < \text{P(in)} < C_T$) we assign an in-manifold result for $\text{P(in)} > 0.5$ and out-of-manifold result for $\text{P(in)} < 0.5$.

At the conclusion of state preparation, we would like to ensure that the molecule resides in one particular sublevel ($\ket{\textsl{i}_J}$ for the experiments presented here). To do that, we require that $\ket{m_g}=\ket{\textsl{i}_J}$ before completing an in-manifold detection. This requirement has the added benefit that preparation can be bypassed after an in-manifold measurement, as indicated in the Experimental Sequence panel of Fig.~\ref{fig:condscheme}.

\subsection{Error budget}

By post-processing our data, we extract information about various error mechanisms affecting our signature manifold SPAM fidelity. Errors due to TR-induced loss are inferred by counting the number of instances in which an out-of-manifold measurement was followed by preparation in a different  $J$  manifold (i.e., taking the difference between the raw and post-selected infidelities plotted in Fig.~\ref{fig:hiFid}). Another possible error mechanism is changes in ion order due to collisions with background gas, which will cause a decrease in detection fidelity due to the change in light shift magnitude for the molecular levels induced by the tightly-focused Raman beams. During SPAM experiments, the ion order is checked and reset as necessary every 15 measurements. We can estimate the infidelity contribution by calculating the fidelity increase when post-selecting away all sets of 15 measurements that are followed by detection of an ion reorder; from this we infer a contribution of $<2.6\times10^{-4}$ for all experimental conditions presented. This is consistent with our observed reorder rate ($\sim$1 reorder/3 minutes from an independent measurement) and the decrease in detection fidelity we observe when purposefully operating in the “wrong” ion order.

Another error mechanism arises from the sub-unity confidence threshold. This error occurs when a sufficiently large number of false-negative outcomes leads to an inaccurate out-of-manifold detection. Assuming the false negative and false positive probabilities used in the Bayesian update are accurate, this error will always be $\leq1-C_T$. To quantify more precisely, we estimate sub-unity $C_T$ error as the average confidence $P(\text{in})$ obtained at the conclusion of Bayesian state detection for out-of-manifold measurements. If this estimate exceeds the post-selected infidelity including ion reorder, the contribution from sub-unity $C_T$ is instead taken to be the post-selected infidelity. Otherwise, the remaining discrepancy reflects residual error due to mechanisms beyond TR-induced loss, sub-unity $C_T$, or ion reorder. We attribute these remaining errors (``other mechanisms") chiefly to signature transition frequency fluctuations, which can increase the actual false negative probability $P(D|\text{in})$. In this case, $P(\text{in})$ will not be faithfully estimated by application of Eq.~\ref{eqn:bayes}. Though we interleave servos of several parameters whose drifts can cause shifts in the signature transition sideband frequency--such as axial OOP motional mode frequency, Raman beam intensities, and micromotion--some drift still occurs. Such errors could be mitigated by more frequent calibration of the signature transition sideband pulse parameters and technical improvements to reduce drifts.

\begin{figure}[htb!]
    \centering
    \includegraphics[width=0.96\linewidth]{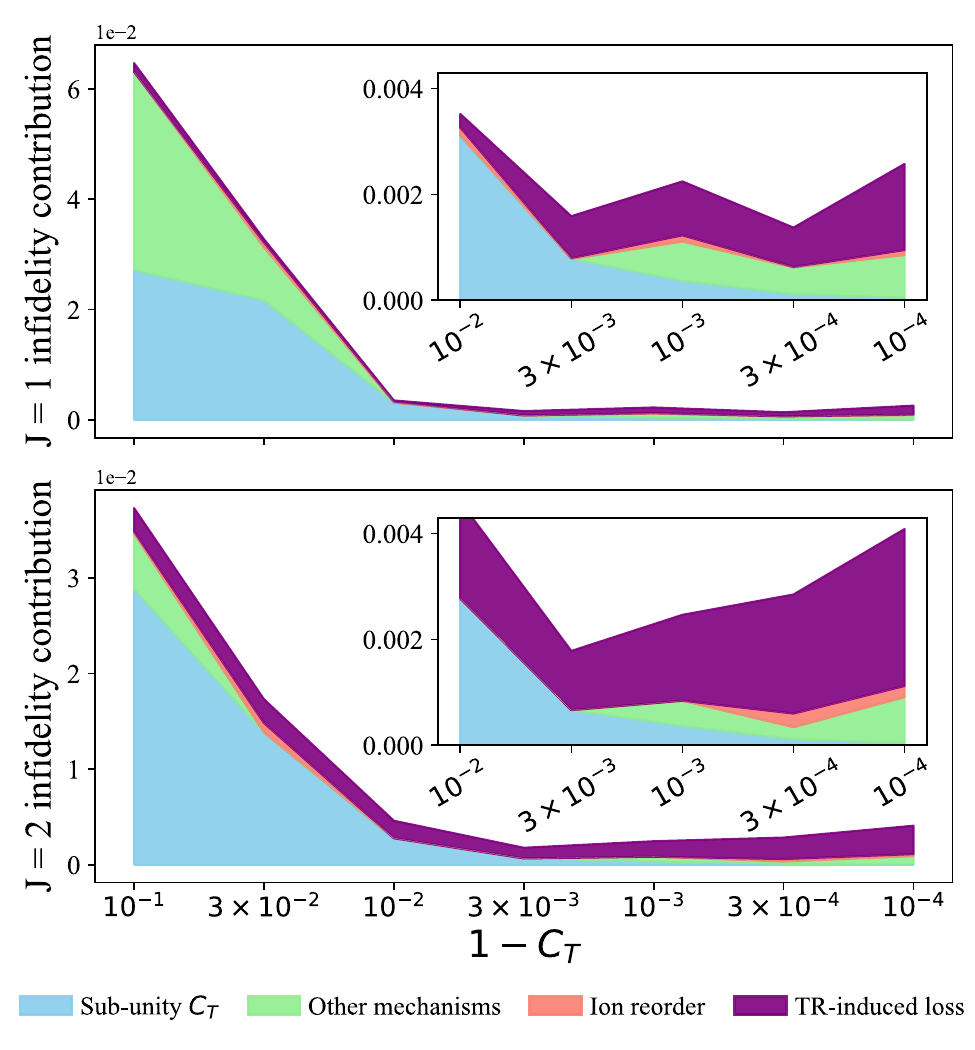}
    \caption{Comparison of infidelity mechanism contributions vs $C_T$. At low $C_T$, most errors are attributable to the sub-unity threshold, while at high $C_T$, TR-induced loss dominates.} 
    \label{fig:ErrMechs}
\end{figure}

\begin{figure}
    \centering
    \includegraphics[width=0.96\linewidth]{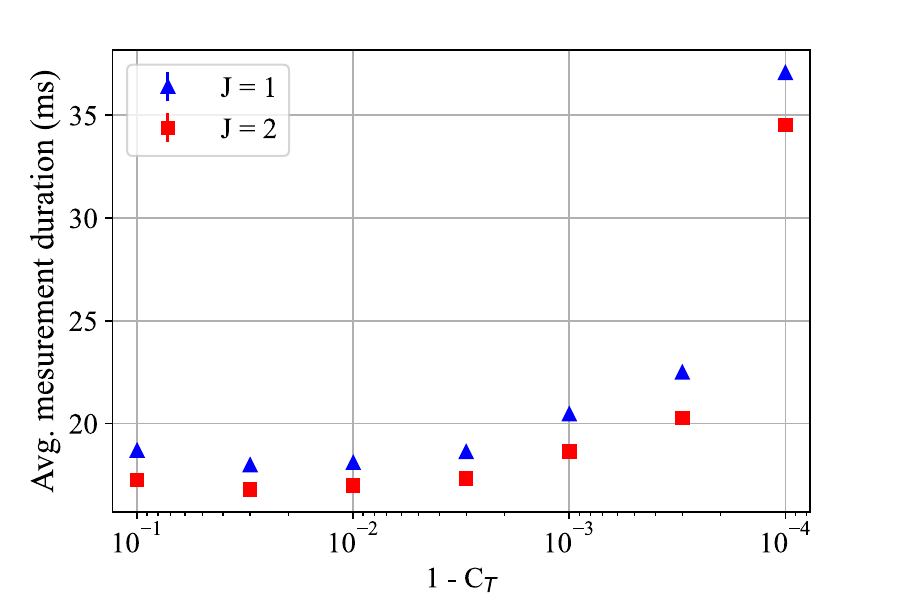}
    \caption{Average measurement duration vs $C_T$. As $C_T$ increases above 0.97, more probes are required to cross threshold, increasing the average measurement duration. Statistical error bars are smaller than the plotted markers.}
    \label{fig:measDur}
\end{figure}

The results of this analysis are displayed in Fig.~\ref{fig:ErrMechs}. As expected, contribution from sub-unity $C_T$ falls off with increasing $C_T$. Meanwhile, the TR-induced loss errors increase with increasing $C_T$ due to the increased average measurement duration. A plot of the average measurement duration for each experiment condition is shown in Fig.~\ref{fig:measDur}.

\bibliographystyle{apsrev4-1}
\bibliography{apssamp}
\clearpage

\section{Supplementary Information}

\subsection{Apparatus}

\begin{figure*}[]
\includegraphics[width=0.8\textwidth]{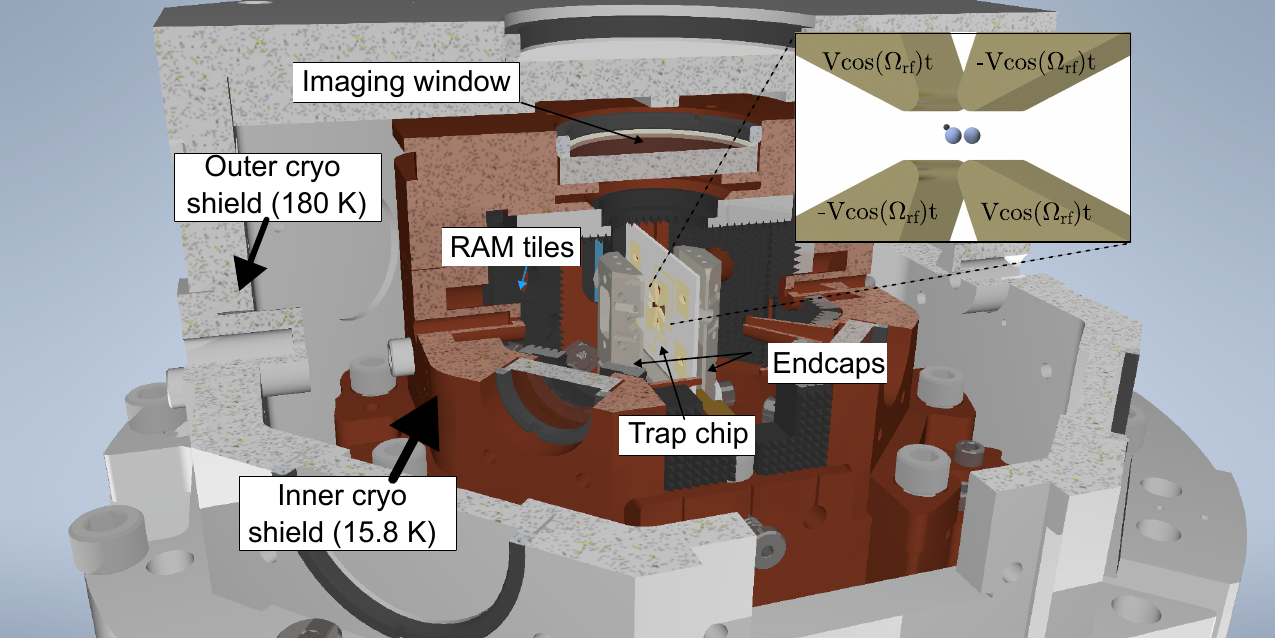}
\caption{\label{fig:appCAD}CAD image presenting a cross section of the experimental setup. Inset shows a zoomed in view of the trapping region.} 
\end{figure*}

A CAD drawing of the experimental setup is presented in Figure~\ref{fig:appCAD}.
The linear radio-frequency (rf) ion trap is housed inside cryogenically cooled radiation shields to reduce thermal radiation and background-gas collisions at the location of the trapped molecular ion. Both shields are cooled by a closed-cycle helium cryostat designed for low-vibration operation and enclosed in an ultra-high vacuum chamber. The inner, gold-plated copper shield reaches a temperature of 15.8~K, while the outer aluminum shield is maintained at 180~K. The cooling power is currently limited by the thermal connection to the cold head but could be improved in the future.

Fused silica and BK7 windows in each shield provide laser and imaging access and partially absorb thermal radiation to prevent leakage to the molecular ion. To ensure ion interaction with leaked radiation is limited in solid angle by windows and apertures, radar-absorbing material (RAM) is included on the inside of the inner shield~\cite{cantatRydbergRAM}. This reduces the possibility of reflections of leaked-in radiation by the inner shield walls.

The trapping potential is provided by a monolithic diamond wafer trap with four rf electrodes driven at 67.63 MHz and a pair of titanium end cap electrodes held at constant electric potential~\cite{chenWheelTrap2017}. We achieve typical secular frequencies near $\{f_z,f_x,f_y\}=\{2.6, 3.8, 4.0\}$ MHz for a single trapped $^{40}\text{Ca}^+$. To achieve sufficient rf voltage amplitudes while minimizing the thermal load on the cold stage of the cryostat, a meander line resonator is located outside of vacuum. The rf signal is routed to the trap using gold-plated stainless steel pins which are thermally anchored at each cryogenic stage using sapphire blocks.

\textsuperscript{40}Ca\Plus~ions are loaded by laser ablation of a CaAl target using a pulsed 1030~nm, 30 $\mu J$ laser source. \textsuperscript{40}CaH\PlusS is formed by reaction with molecular hydrogen. A molecular beam which enters through a hole in the end cap electrode can be used to load other molecular ion species for future experiments.

\subsection{Rotational state lifetime limitations}

The inferred BBR environment temperatures reported in Table~\ref{tab:table1} are higher than the measured inner shield temperature of 15.8 K. In this section, we present an experimental investigation of potential mechanisms that could account for this discrepancy.

\begin{itemize}
    \item \textbf{Trap electrode temperature}: Under normal operating conditions, about 1 W of rf power is dissipated on the inner shield. Though expected thermal gradients across the diamond trap chip are minimal, localized defects or poor thermal contact could raise electrode temperatures, thus increasing the TR experienced by the ion. With no rf power dissipated on the inner shield we observe at temperature of 11.9 K which is lower then the 15.8 K temperature observed during normal operation. We have performed a continuous molecular rotational state tracking experiment with trap rf power elevated by a factor of 1.6. The results are presented in Table~\ref{tab:table3}. At elevate rf power we observe an increase in the inner shield temperature from 15.8 K to 18.5 K. We have not observed a statistically significant deviation from the reported values at lower rf power.
    \begin{table}[htbp]
    \caption{\label{tab:table3}Rotational lifetimes and inferred BBR temperatures obtained by tracking the molecular rotational state for 43 minutes at elevated rf power. For comparison, the right most column represents inferred BBR temperature when operating at normal rf power (Table \ref{tab:table1}).}%
\renewcommand{\arraystretch}{1.25}
\begin{ruledtabular}
\begin{tabular}{cccc}
\textrm{$J$}&
\textrm{Lifetime (s)}&
\textrm{BBR temp.} &
\textrm{BBR temp.,normal rf power}\\ 
\colrule
1 &  $16.9\pm2.6$& $44.7^{+6.9}_{-5.3}$& $42.0^{+4.3}_{-3.7}$\\
2 &  $9.8\pm1.5$& $34.7^{+5.4}_{-4.2}$& $33.6^{+3.2}_{-2.9}$\\
3 &  $5.7\pm1.1$& $29.7^{+7.2}_{-5.5}$& $30.6^{+4.1}_{-3.6}$\\
\end{tabular}
\end{ruledtabular}
\end{table}

    \item \textbf{Thermal radiation leakage}: TR from warm surfaces can reach the ion via BK7 and fused silica windows and apertures. BK7's absorption at 285 GHz was experimentally measured by comparing the Rabi rate of the molecular$J=0\leftrightarrow1$ transition with and without an additional BK7 window, which was positioned in the microwave beam path outside of vacuum. For 285 GHz we have found an absorption coefficient $\alpha=1.61$ (cm$^{-1}$). We scaled the corresponding $\alpha$ for 570 GHz using quadratic power law~\cite{naftaly2005terahertz}. The corrected lifetimes were calculated by scaling the photon energy density according to the relative solid angle of apertures and windows, taking into account the attenuation at the microwave regime. For fused silica windows we assume no attenuation. We compare the expected lifetimes in a BBR environment at 15.8 K with and without such corrections in Table~\ref{tab:table2}.

\begin{table}[htbp]
\caption{\label{tab:table2}%
Calculated rotational life times of CaH\Plus interacting with BBR environment at 15.8 K, including and excluding leakage of hotter TR through apertures and windows.
}
\renewcommand{\arraystretch}{1.15}
\begin{ruledtabular}
\begin{tabular}{ccc}
\textrm{$J$}&
\textrm{Lifetime (s)}
&
\textrm{Lifetime, leakage included (s)}\\
\colrule
1 & $73.4^{+5.6}_{-5}$& $66.6^{+5.0}_{-4.5}$\\
2 & $24.5^{+1.8}_{-1.6}$& $23.0^{+1.7}_{-1.5}$\\
3 & $9.3^{+0.7}_{-0.6}$& $8.8^{+0.7}_{-0.6}$\\
\end{tabular}
\end{ruledtabular}
\end{table}

    \item \textbf{Background gas collisions}:
    
    Collisions with background gas can cause transitions between rotational levels. To empirically investigate the effect of background gas collisions, we have performed state tracking experiments at various levels of background pressure~\cite{Chou_tracking}. The collision rate is proportional to the ion crystal reorder rate which was measured at each pressure level. At each reorder rate we performed a molecular state tracking experiment and extracted the rotational lifetime per $J$ state. The lifetime vs. reorder rate data was then fitted to a linear curve using parametric bootstrapping. We observe a slope of $-167\pm100$ ($-87\pm32$) s/s$^{-1}$ and an offset of 21$\pm$2 (10$\pm$1)~s for $J$=1(2). The point with the lowest reorder rate of each dataset represents our operating pressure and corresponds to the state tracking data presented in Table~\ref{tab:table1}. The observed nonzero correlation between rotational lifetimes and reorder rate indicates possible effect of background gas collisions at high pressures. However, at our operating pressure, the experimentally observed lifetimes agree within reported uncertainties with the extrapolated lifetimes at zero pressure. This conclusion relies on the linear dependence between background pressure and the Langevin collision rate~\cite{wineland1998bible,koelemeij2007blackbody}. We therefore conclude that within the uncertainty of our measurements, background gas collisions do not affect the rotational lifetimes reported in Table \ref{tab:table1}.
    

\begin{figure}[htbp]
\includegraphics[width=0.49\textwidth]{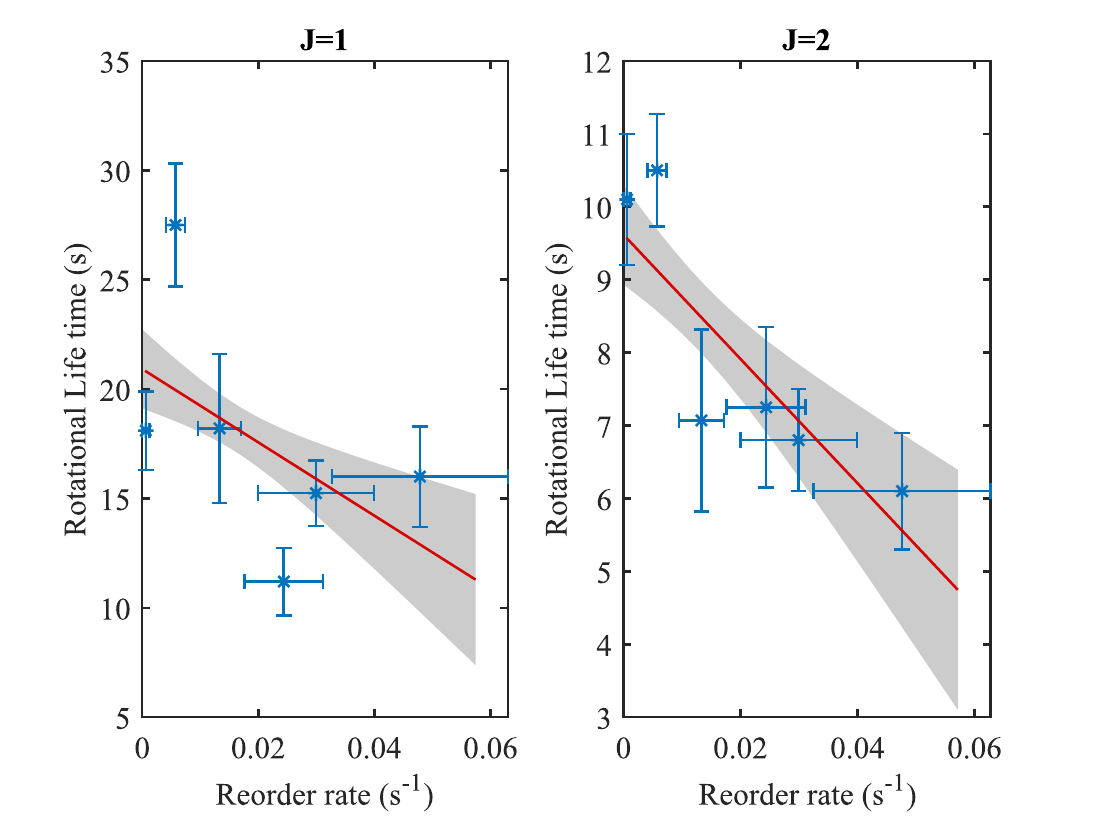}
\caption{\label{fig:Collisions}Observed correlation between rotational life time and the two ion crystal reorder rate. Linear fit and 1 STD fit uncertainties are presented by a red line and gray area.} 
\end{figure}

    \item \textbf{Laser beams induced effects}: 
    During molecular state tracking and SPAM measurements, the molecule is addressed by various laser beams including the atomic cooling and readout laser beams at 397, 854, 866 and 729~nm and Raman beams at 1064~nm. A molecular state change might be induced, for example, by an off-resonant electronic transition driven by 397~nm light. The atomic and Raman laser beams might also induce a thermal effect, locally heating a surface with solid angle to the ion. Such heating might result in hotter BBR interacting with the ion. We have observed correlations between motional mode frequency drifts and Raman beam intensities which we attribute to thermal effects. 
    For further investigation of these effects we have performed an additional experiment directly measuring the molecule's rotational state lifetime. Here a wait time was added between preparation and measurement during which the cooling light at 397~nm at reduced intensity and re-pumping laser light at 866 and 854 nm were on. Before detection, ion order and the ion crystallization were verified. The rotational lifetime is extracted from the observed decay in rotational state population as a function of the wait time duration (Figure \ref{fig:WT}). The results are shown in Table~\ref{tab:table4}. 
    The inferred BBR temperatures agree better with the inner shield temperature sensor reading of 15.8~K indicating the possibility of off resonantly exciting the molecule or a thermal effect occurring during preparation or detection. 

\begin{figure}[htbp]
\includegraphics[width=0.47\textwidth]{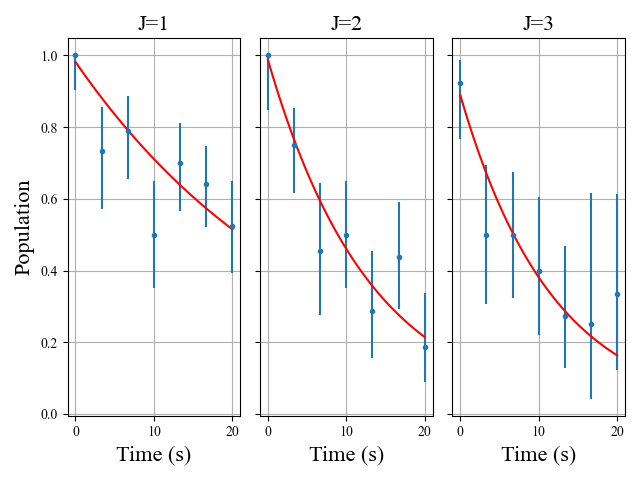}
\caption{\label{fig:WT} Survival probability of the molecule in state $J$ within the signature manifold as function of time between preparation and measurement. Fitted line to a single exponent is shown in red.} 
\end{figure}

    \begin{table}[htbp]
    \caption{\label{tab:table4}Rotational lifetimes as obtained by exponential fits of the experimental data shown at Figure \ref{fig:WT}. Inferred BBR temperatures are presented as well.}%
\renewcommand{\arraystretch}{1.25}
\begin{ruledtabular}
\begin{tabular}{cccc}
\textrm{$J$}&
\textrm{Lifetime (s)}&
\textrm{BBR temp. (K)}\\
\colrule
1 &  $29.3\pm4.6$& $29.2^{+4.2}_{-3.2}$\\
2 &  $16.5\pm3.2$& $22.7^{+4.6}_{-3.5}$\\
3 &  $8.0\pm1.1$& $20.2^{+4.4}_{-4.1}$\\
\end{tabular}
\end{ruledtabular}
\end{table}

    \item \textbf{Geometrical effects}: The trap electrode separation (0.5 mm) is on the order of the wavelength of relevant radiation (0.4-1.1 mm for transitions affecting the lifetimes of $J=1,2$ sublevels), implying that near-field effects may shape local TR spectrum~\cite{Chou_tracking, BBR_considerations}. It may contribute to the observed reduced lifetimes.
    
\end{itemize}

\end{document}